\documentclass[lettersize,journal]{IEEEtran}
\IEEEoverridecommandlockouts
\usepackage{cite}
\usepackage{multirow}
\usepackage{amsmath,amssymb,amsfonts}
\usepackage{algorithmic}
\usepackage{algorithm}
\usepackage{graphicx}
\usepackage{array}
\usepackage[caption=false,font=normalsize,labelfont=sf,textfont=sf]{subfig}
\usepackage{textcomp}
\usepackage{stfloats}
\usepackage{url}
\usepackage{verbatim}
\usepackage{xcolor}
\usepackage{balance}
\def\BibTeX{{\rm B\kern-.05em{\sc i\kern-.025em b}\kern-.08em
    T\kern-.1667em\lower.7ex\hbox{E}\kern-.125emX}}
    \usepackage{balance}
\begin{document}

\title{New Collision Free Balanced Frequency Hopping Sequence Sets: Construction and Analysis\\

}
\author{\IEEEauthorblockN{M Kavi Priya  ~~ and ~~ K Giridhar}\\
\IEEEauthorblockA{\textit{Department of Electrical Engineering } \\
\textit{Indian Institute of Technology Madras}\\
Chennai, India \\
}
}

\maketitle

\begin{abstract}
Collision-free Frequency Hopping Sequences (FHS) are crucial for maintaining the throughput of Frequency Hopping Multiple Access (FHMA) communication systems. When multiple FHS deployed in the same geographical area hop into the same frequency spot or location, it can result in multiple access interference (MAI), leading to performance degradation of the co-habiting FHMA systems. Orthogonal FHS (O-FHS) could be used to avoid such collisions even while maintaining the pseudo-random hopping property. We propose a new Collision-Free Balanced Frequency Hopping Sequence (CFB-FHS) set based on $GF(p)$. The proposed algorithm generates a significantly longer O-FHS and ensures full-band hopping and uniform usage of all the available frequency spots, even while preserving the pseudo-random nature of the sequences. The Hamming correlation and these related properties of the proposed sequence set are analyzed and compared with the existing sequence sets. The length of this CFB-FHS can be chosen based on the application requirements, and once selected, the length remains fixed irrespective of the family size of the sequence set.
\end{abstract}

\section{Introduction}
Frequency Hopping (FH) Spread Spectrum is a communication technique widely used in many applications due to its secure nature and anti-jamming properties. Frequency Hopping Sequences (FHS) are used in FH systems to specify the frequency sub-band or spot to transmit the data. It is essential that FHS is random, and the longer the period of these sequences, the more secure the system is. FHS construction based on $m$-sequences has been proposed in \cite{b2}, and it also describes a Hamming correlation approach to find the randomness of any sequence. The authors also derived the lower bound on the Hamming correlation of two sequences for an optimal FHS. In \cite{b3}, a lower bound on Hamming Correlation for the FHS set where the family size is greater than two has been derived. The FHS set that achieves equality in the Peng-Fan bound from \cite{b3} is considered to be an optimal FHS set. 

When multiple FH systems have to co-exist in the same area, it is called Frequency Hopping Multiple Access (FHMA). FHS with orthogonality is used in FHMA applications to prevent Multiple Access Interference (MAI) caused when two or more FHS simultaneously visit the same frequency spot. Such sequences are referred to as orthogonal FHS (O-FHS) or No-hit zone FHS (NHZ-FHS). The NHZ-FHS were proposed in \cite{b4} where the sequences have zero correlation when the relative time delay between the FHS is within the NHZ region. This work also described the construction of an NHZ-FHS set and gives the bound on zero correlation zone based on the number of available frequency sub-bands, $M$, and the family size, $q$. Recent works \cite{b5} - \cite{b7} discuss the construction of NHZ-FHS sets with reduced Hamming correlation outside the NHZ, and the bounds on maximum Hamming correlation outside the correlation zone were derived.

Existing literature on NHZ-FHS is rather limited, and the O-FHS sets with longer sequence period could be very useful for future secure communication systems. Our paper introduces a novel Collision-Free Balanced Frequency Hopping Sequence (CFB-FHS) set that is constructed based on finite fields, $GF(p)$, where every sequence in the set is orthogonal to each other. The proposed CFB-FHS set can achieve a longer sequence period irrespective of the family size, $q$. Additionally, our sequences do a full-band hopping, use all the available frequency spots equally and balance every sequence in such a way that their pseudo-randomness is retained. The forthcoming sections explain the construction of this CFB-FHS set and present the complete analysis of the generated sequence set and its properties.

\section{System Model and Notations}
Consider a communication system where multiple FH subsystems have to co-exist in the same geographic area. Each subsystem has to hop orthogonally to transmit in different frequency spots. Let the communication system have $q$ subsystems transmitting signals in a frame structure with multiple hops in each frame. All the subsystems hop at the same frequency rate and are assumed to be frame synchronized to 1-PPS using the Global Positioning System (GPS) or any other equivalent global time reference. It is assumed that any relative time delay between the nodes in the network is minimal compared to the dwell time of the system; i.e., the relative time delays are well within one hop duration. This section further briefly explains some preliminary notations required for the forthcoming sections.

Given a prime number $p$ and a positive integer $l$, consider $l$ to be the degree of the primitive polynomial used in constructing a linearly generated maximal-length sequence (also known as $m$-sequence). An $m$-sequence $S=\{s(0),s(1),...,s(n-1)\}$ over $GF(p)$ has sequence period, $n = p^l - 1$. The $m$-sequences and their properties are discussed in detail in \cite{b2} and \cite{b1}. Let $b$ be a positive integer such that $S_b(j)$ is a $b$-tuple word from  $S(j)$; i.e., $S_b(j) = \{ s(jb),s(jb+1),...,s(jb+b-1)\}$ of length $b$, where $ 0 \leq j < n'$ and $n' = \lfloor \frac{n}{b}  \rfloor $. Let $F = \{ f_0,f_1,...,f_{M-1}\}$ be the set of all available frequency spots, where the number of frequency spots is given by $M = p^b$. Let $L = \{l(j)\}$ of length $n'$ over $F$ have a mapping from every $b$-tuple $S_b(j)$ given by
\begin{equation}
l(j) = S_b(j) = \sum_{i = 0}^{b-1}  s(jb+i) p^i , \; \; \; \;   0 \leq j < n'
\label{FHS transform}
\end{equation}

\section{Construction of CFB-FHS sequence}
This section presents our algorithm for generating a set $\mathcal{C}$ consisting of $q$ sequences that are orthogonal to each other and suitable for synchronous FHMA systems. The construction of CFB-FHS set $\mathcal{C}(M,n',q)$  consists of the following two steps. 

\underline{Step 1}: We generate a sequence set $\mathcal{K}$ of size $q$. We start with an $m$-sequence, $S$, over $GF(p)$ of length $n$. The sequence set $\mathcal{K} = \{ L^0,L^1,...,L^{q-1}\}$ over $F$, consists of $q$ sequences, where $L^a$ denotes the $a^{th}$ sequence of $\mathcal{K}$, $0 \leq a \leq q-1$. Each $L^a$ sequence of length $n'$ represents a one-to-one mapping of the maximal-length sequence $S^a$ of length $n$. The sequence $S^a$ is $S^a = \{ s(a\tau )_n,s(a\tau +1 )_n,...,s(a\tau +n-1)_n \}$, where $\tau $ can be any prime number such that $0 \leq \tau < n$ and $0 \leq a < q$. Then, $b$-tuple word from $S^a$ can be expressed as $S^a_b(j)= \{ s(a\tau +j )_n,s(a\tau +j+1 )_n,...,s(a\tau +j+b-1)_n \}$. By considering every $b$-tuple word $S^a_b(j)$ and mapping them to $F$, we obtain the sequence $L^a = \{ l^a(j)\}$ with $0 \leq j \le n'$, which is given by:
\begin{equation}
l^a(j) = S^a_b(j)= \sum_{i=0}^{b-1} s(a\tau + jb+ i)_n  \; p^i
\label{seq set gen}
\end{equation}
where $(.)_n$ represents the modulo-$n$ operation. Note that the generated set $\mathcal{K}$ has $q$ sequences that are not orthogonal to each other. 

\underline{Step 2}: From the sequence set $\mathcal{K}$, the orthogonal FHS is generated using CFB-FHS scheme, following Algorithm 1. (i)  The variables $\alpha$ and $\alpha'$ keep track of the number of operations performed on all $q$ sequences in $\mathcal{K}$ and the number of times each frequency slot is used, respectively. (ii) For every $i\in\{1,2,..., n'\}$, $\beta$ has the $i^{th}$ entry of all the $q$ sequences. (iii) For every repetition of the frequency sub-bands, $F = \{ f_0,f_1,...,f_{M-1}\}$, in $\beta$, replace the $i^{th}$ entry of the less operated sequence among the repeated ones with one of the frequency slot not used in $\beta$. (iv) update $\alpha$ and $\alpha'$. 

By using the above steps, we obtain an orthogonal FHS set $\mathcal{C}$ that contains $q$ sequences, each with a length of $n'$. The properties of this sequence set are analyzed in Section IV.

\section{Properties and Analysis of CFB-FHS Set }

This section discusses the various properties of the proposed CFB-FHS set and provides relevant results. The performance of an FHS depends on parameters such as the number of available frequency spots, $M$, the sequence period, $n'$, the size of the family, $q$, and the Hamming correlation of the sequence. The performance of our CFB-FHS set $\mathcal{C}(M,n',q)$ is compared with existing sequences, namely, Strong NHZ-FHS \cite{b5} and Multi-Level FHS \cite{b6}, which are referred to as SNHZ-FHS and ML FHS, respectively, throughout this paper.

\subsection{Theoretical Bound for CFB-FHS Set}
Based on the available number of frequency slots, the upper bound for the number of orthogonal users can be fixed. The number of sequences, $q$, in the set $\mathcal{C}$ that represents the maximum number of users in the FHMA system cannot exceed the available frequencies $M$ to ensure a collision-free communication system. Hence
\begin{equation}
1 \leq q \leq M, \;\;\;\;\; with \;\; M = p^b
\label{q constraint}
\end{equation}
The above constraint for CFB-FHS set follows from the fact that if $q$ exceeds the available number of spots $M$, then collision is inevitable.

\subsection{Uniform Frequency Usage}
\begin{figure}[h]
\centerline{\includegraphics[scale=0.21]{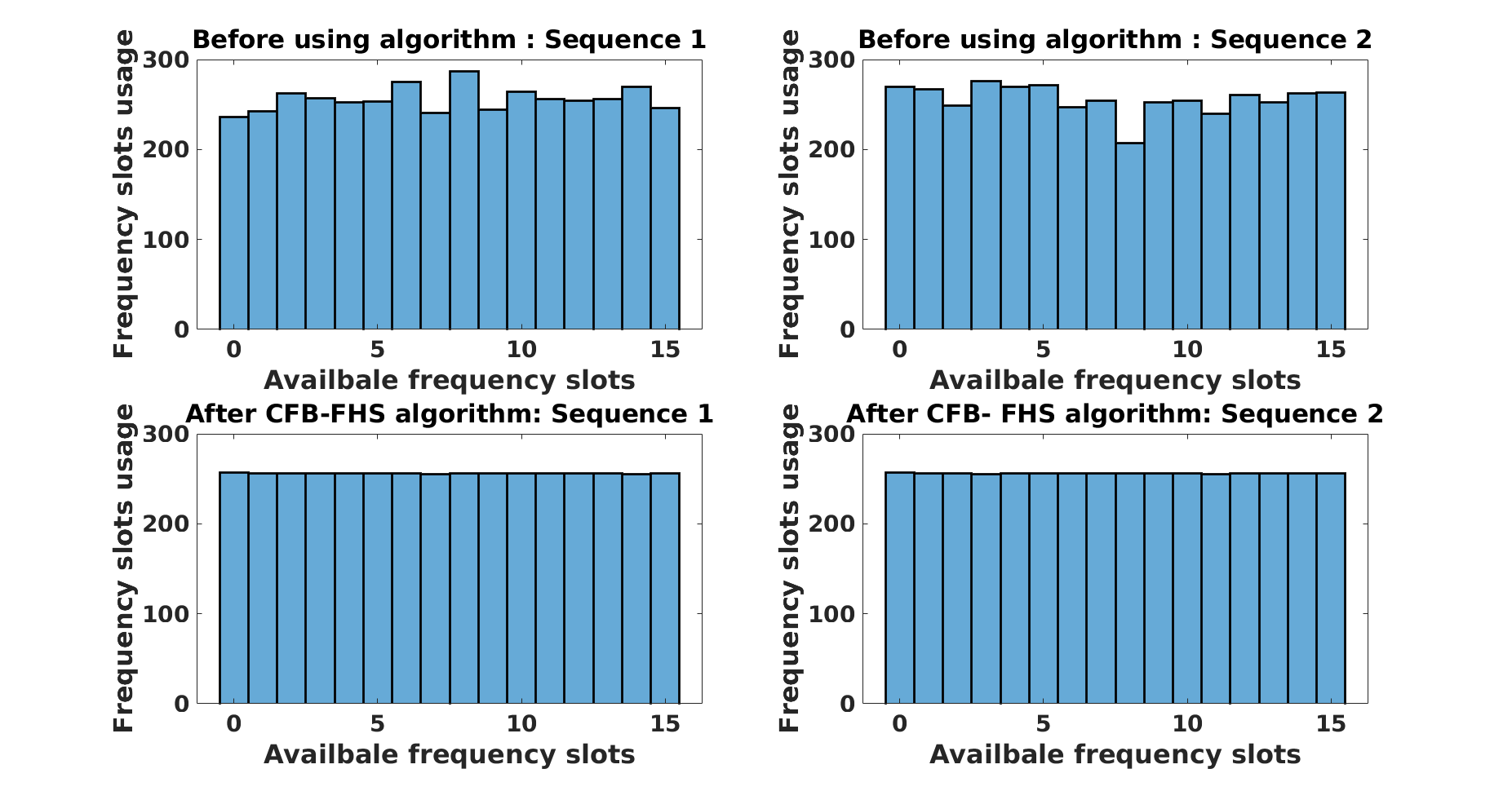}}
\caption{Histogram of the frequency spots usage of sequences from set $\mathcal{K}$  and $\mathcal{C}$ for $M$ = 16, $q$ = 5, $l$ = 14 and length = 4095}
\label{bef_aftr_his_comp}
\end{figure}

This section analyses the usage of every frequency bands $F = \{ f_0,f_1,...,f_{M-1}\}$ by every users. When all the users use all the available frequency bands equally over a period of time, we can ensure that fairness is maintained among the users. The proposed CFB-FHS set ensures full-band hopping even while equally using all the available $M$ frequency spots. The sequences in the set $\mathcal{K}$ after Step 1 do not use the available frequency spots equally, and after the CFB-FHS algorithm, the constructed sequences in set $\mathcal{C}$ use all the spots an equal number of times. The histogram of two sequences from set $\mathcal{K}$, referred to as sequences before using the algorithm, is shown at the top of Fig.~\ref{bef_aftr_his_comp}, and the histogram of the same two sequences from CFB-FHS set $\mathcal{C}$ is shown at the bottom of Fig.~\ref{bef_aftr_his_comp}. As we can see from the figure, sequences from set $\mathcal{C}$ use every frequency spot nearly equally.

\begin{figure}[h]
\centerline{\includegraphics[width=9cm,height=4.5cm]{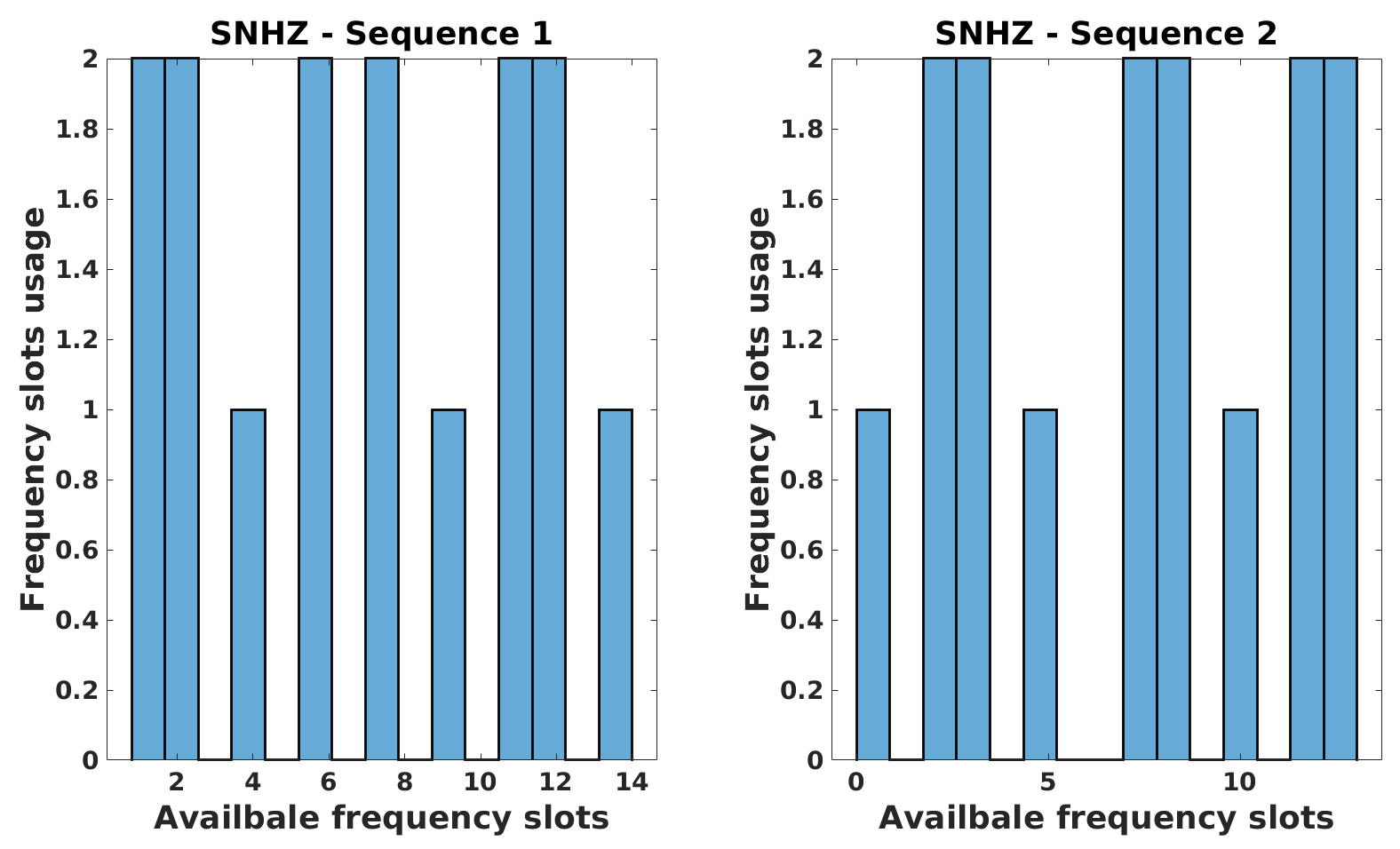}} 

\caption{Histogram of the frequency usage of SNHZ-FHS set, $M$ = 15, $q$ = 5, length = 15}
\label{snhz_hist}
\end{figure}

\begin{figure}[h]
\centerline{\includegraphics[width=9.5cm,height=4.5cm]{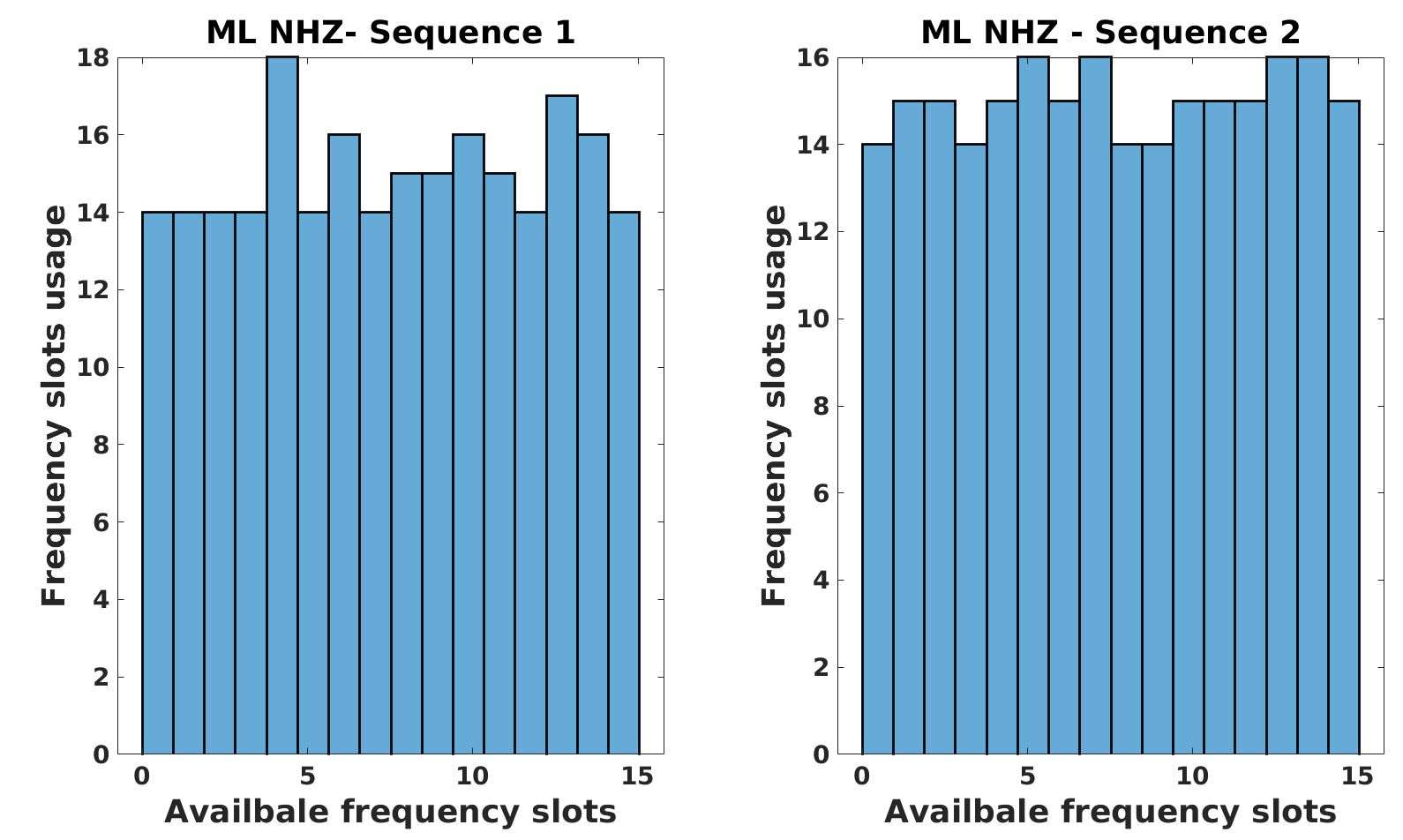}} 
\caption{Histogram of the frequency usage of ML FHS set, $M$ = 16, $q$ = 5, length = 240}
\label{ml_nhz_hist}
\end{figure}

Existing methods in literature, such as \cite{b5} and \cite{b6}, have histograms for the usage of frequency spots as shown in Fig.~\ref{snhz_hist} and Fig.~\ref{ml_nhz_hist}. These histograms illustrate that the available frequencies are not equally used in the existing methods. In contrast, our proposed method generates FHS that does full-band hopping and uses all the available frequency spots equally.

\subsection{Fairness of CFB-FHS Procedure}

Every sequence, $C^a$,  in the CFB-FHS set $\mathcal{C} = \{ C^0, C^1,..., C^{q-1}\}$ undergoes $d_a$ number of operations by the algorithm when getting transformed from set $\mathcal{K}$ to $\mathcal{C}$, $0 \leq a < q$. The proposed CFB-FHS treats all the sequences fairly, ensuring that the random nature of frequency spot usage is still maintained. For a given $q$, the average number of operations done per sequence in the set is $\bar{d} = \frac{1}{q} \sum_{i = 0}^{q-1}  d_i$. 

\begin{figure}[h]
\centerline{\includegraphics[scale=0.158]{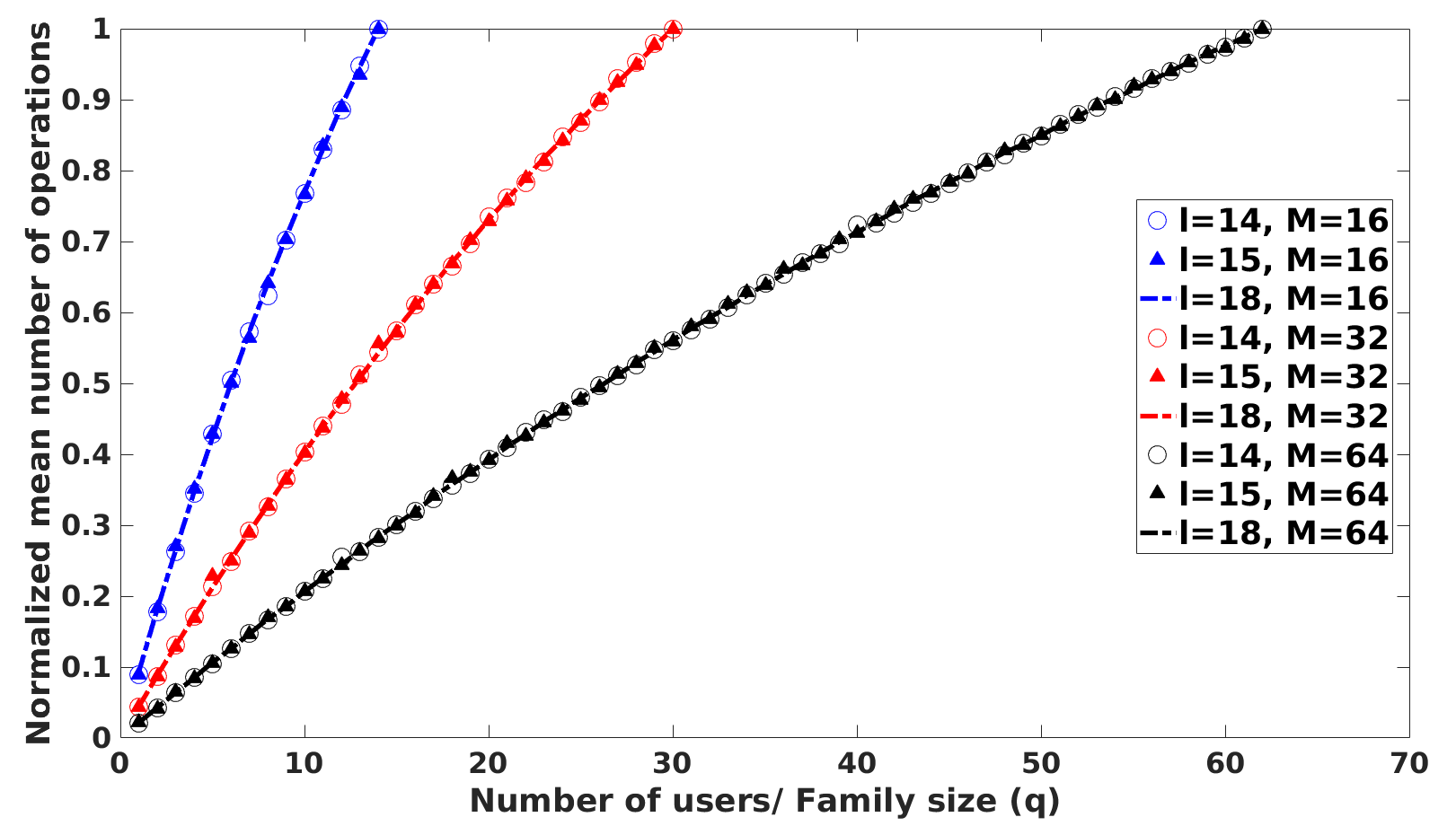}}
\caption{The plot of $\bar{D}$ for cases $l = 14,15,18$ and $M = 16,32,64$ vs $q$}
\label{swap trend}
\end{figure}

This $\bar{d}$ is analyzed for various sets generated for difference $l$ and $M$ values. For a fixed $l$ and $M$ values, the $\bar{d}_q$ is calculated for every possible value of $q$ such that $0< q \leq M$. The vector $D = \{\bar{d}_0, \bar{d}_1, ... , \bar{d}_M\} $ is normalized to get the mean operation vector, $\bar{D}$. This normalized vector $\bar{D}$ is calculated for every combination of $l = 14,15$ and $18$ and the corresponding $M = 16,32$ and $64$, and plotted in Fig.~\ref{swap trend}.

 It can be seen from this figure that $\bar{D}$ follows a linear trend, and we use a least square fitting on the data using MATLAB. The data points follow the linear model, $f(x) = h_1x + h_2$, where $h_1$ and $h_2$ are the slope and intercept respectively. Also, $h_2$ approximately takes the same value for all cases whose mean value is $h_2 = 0.0605$. The $h_1$ values for observed scenarios are tabulated in Table~\ref{tab_swap_trend}. From this table, it can be inferred that the $h_1$ values are independent of the degree of the primitive polynomial, $l$, and depends only on the available frequency spots $M$. The relation between $h_1$ and the number of frequency locations can be expressed as,
\begin{equation}
h_1 \approx \frac{1}{M}
\label{slope}
\end{equation}
From the above observations, it is clear that as the number of sequences $q$ in the set increases, the number of computations done also increases linearly.

\begin{figure}[h]
\centerline{\includegraphics[scale=0.158]{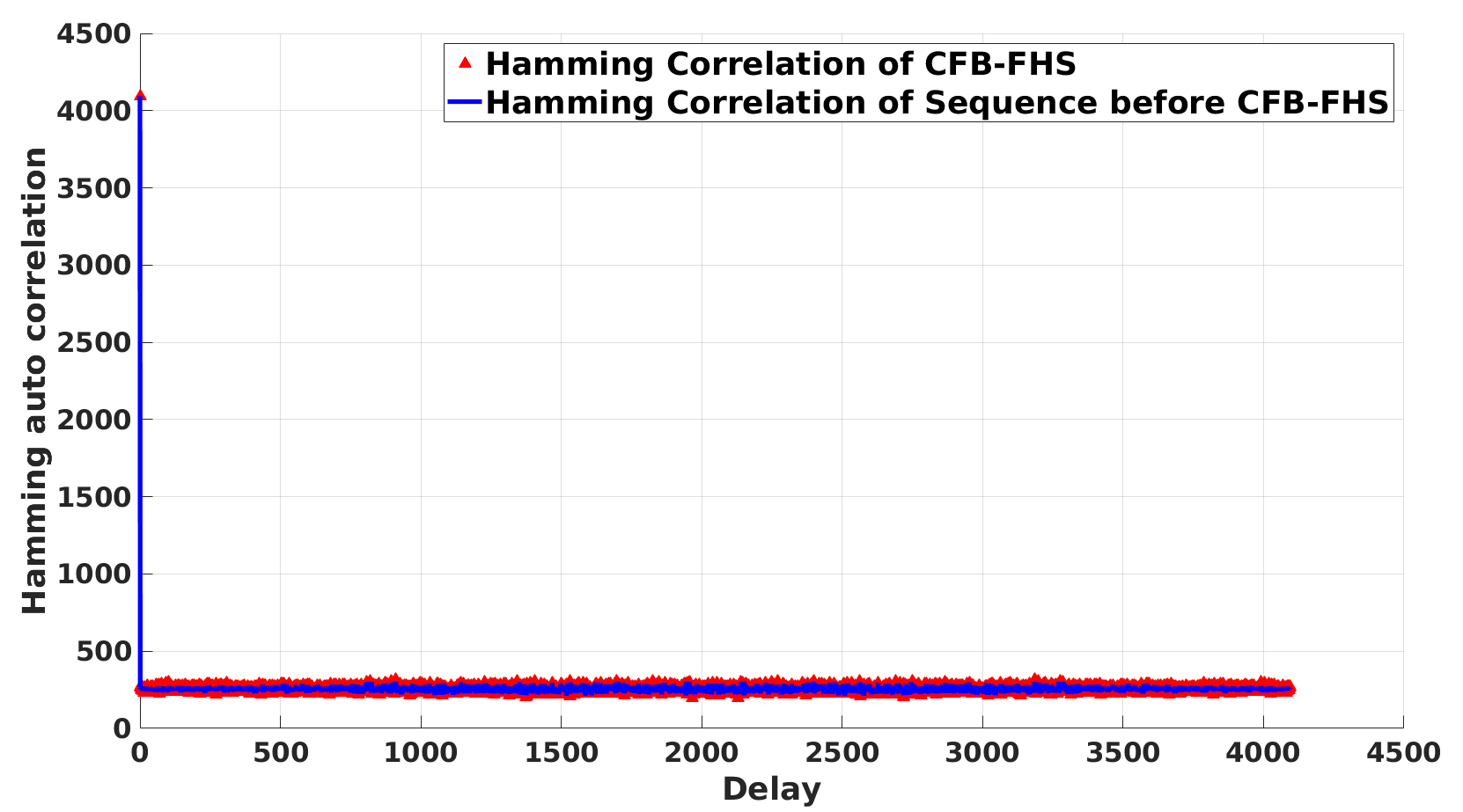}}
\caption{Hamming auto correlation comparison of a sequence before and after applying CFB-FHS algorithm}
\label{Hamming_corr_comp_v2}
\end{figure}

\begin{figure}[h]
\centerline{\includegraphics[scale=0.158]{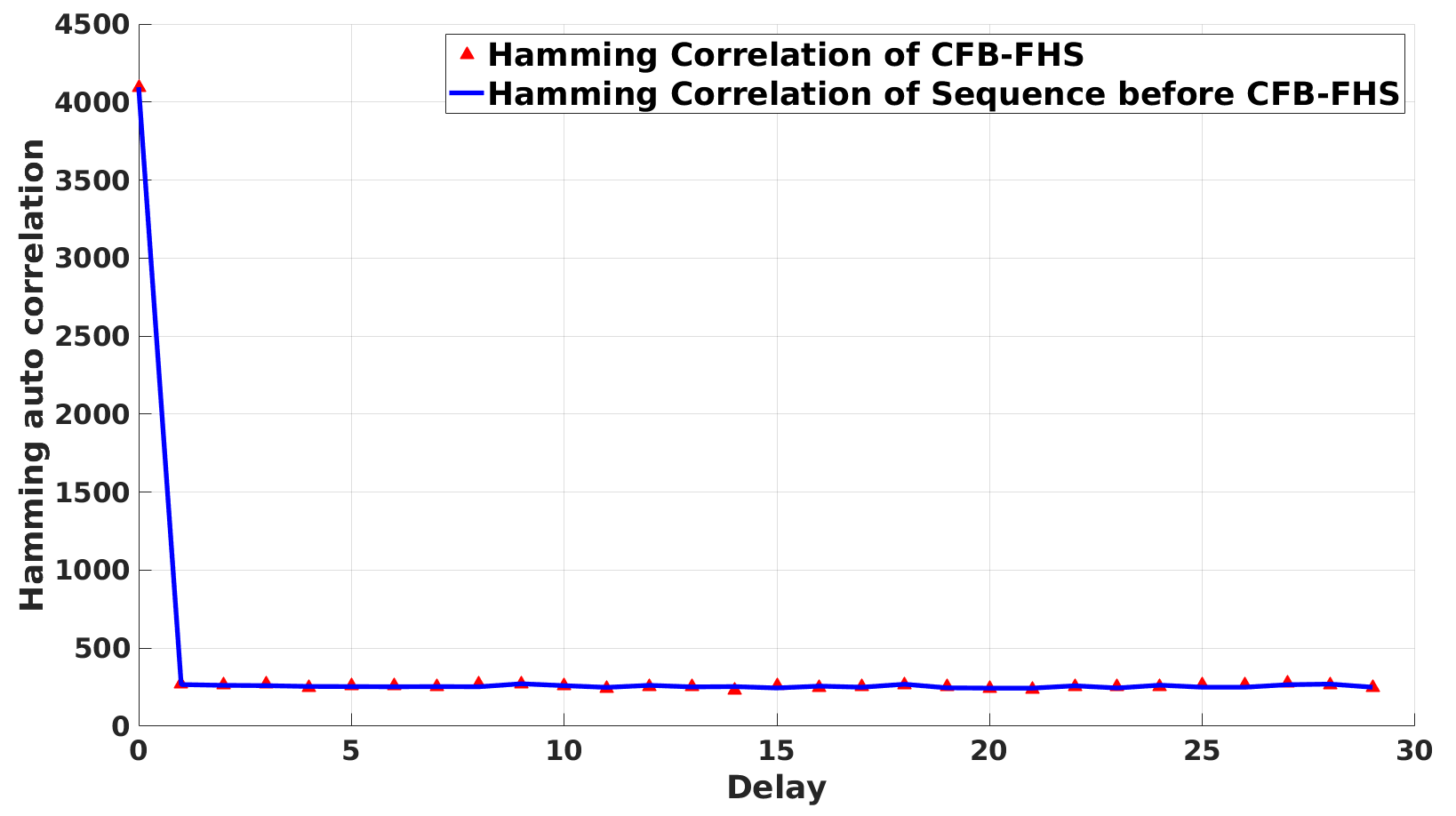}}
\caption{Hamming auto correlation comparison of a sequence before and after applying CFB-FHS algorithm}
\label{Hamming_corr_zoom}
\end{figure}

The generated CFB-FHS set retains the random nature of the sequences even after undergoing a number of operations by the algorithm. The random nature of a sequence is measured by its Hamming auto-correlation property. The Hamming auto-correlation plot of a sequence from the set $\mathcal{K}$ and the same sequence transformed to $\mathcal{C}$ is shown in Fig.~\ref{Hamming_corr_comp_v2}. Here, we can see that the maximum value of the Hamming auto-correlation is $n' = 4095$, which is the length of the sequence for both sequences. It can also be seen that both these sequences possess similar correlation properties for all delay values. Since Fig.~\ref{Hamming_corr_comp_v2} has delay values up to $4095$, the zoomed in version of Fig.~\ref{Hamming_corr_comp_v2}, (i.e., the Hamming auto-correlation plot for the same two sequences for delay value upto $30$ alone) is shown in Fig.~\ref{Hamming_corr_zoom} for better clarity. In this figure, we can see the auto-correlation values exactly coincide for the sequences from $\mathcal{K}$ and $\mathcal{C}$. This result indicates that the proposed CFB-FHS retains the random nature of the sequence.

\begin{table}[htbp]
\caption{Values of slope($h_1$) of the computational count linear model}
\begin{center}
\begin{tabular}{|p{1.5cm}|p{1.5cm}|p{1.5cm}|}
\hline
\textbf{Degree of Polynomial, $l$} & \textbf{Number of frequency spots, $M$} & \textbf{Slope of the computational count} \\ 
\hline
14 & 16 & 0.06955 \\
\hline
15 & 16 & 0.06897 \\
\hline
18 & 16&0.06949\\

\hline
14 & 32 & 0.03278 \\
\hline
15 & 32 & 0.03253 \\
\hline
18 & 32&0.03278\\

\hline
14 & 64 & 0.01586 \\
\hline
15 & 64 & 0.0161\\
\hline
18 & 64 &0.01593\\
\hline
\end{tabular}
\label{tab_swap_trend}
\end{center}
\vspace{-6pt}
\end{table}

\subsection{Sequence Period}
The period of the $m$-sequence $S$ over $GF(p)$ is $n = p^l -1$ where $l$ is the degree of the primitive polynomial used. It can be seen that the length of the sequence depends on the $p$ and $l$ values. In all cases discussed in this paper, we use $p = 2$ since we are working with binary $m$-sequence. The degree of the polynomial $l$ can be fixed according to the required period of the sequence, which varies for different applications.

When the frequency hopping sequence set $\mathcal{K}$ is constructed from the $m$-sequence, the length of every sequence $L^a$ reduces to $n'=\lfloor \frac{n}{b}  \rfloor $, where $b$ is the number of tuples used in every word used. The $b$ is chosen based on the available frequency spots for frequency hopping. The algorithm generating the sequence set $\mathcal{C}$ does not alter the length of the sequences from set $\mathcal{K}$, whatever be the family size $q$ that is chosen. It can be seen that the period $n'$ of a sequence from $\mathcal{C}$ is independent of the number of users $q$, given that the constraint in \eqref{q constraint} holds and a longer period can be chosen by choosing appropriate $l$ for a fixed $M$ irrespective of the number of users, $q$. 

In Table~\ref{length_comp}, the length of sequences generated by existing methods, \cite{b5} and\cite{b6}, are compared with our proposed CFB-FHS algorithm. The table shows that regardless of the number of users in the system, our algorithm can generate very long sequences. Also, by selecting an appropriate value for $l$, the length of the sequence can be increased for any given application. Thus, the CFB-FHS algorithm can accommodate any number of users with longer sequence periods, and the only trade-off is choosing a larger primitive polynomial.

\subsection{Correlation Property of the Sequences}
The Hamming correlation \cite{b2} of the two sequences $C^u$ and $C^v$ each of length $n'$  for an integer delay $\alpha$ is,
\begin{equation}
G_{uv}(\alpha) = \sum_{i=0}^{n'} g(C^u_i,C^v_{i+\alpha})
\label{Ham_Cor}
\end{equation}
where $g(x,y) = 1$ for $ x=y$ denotes the collision between the two sequences, and $g(x,y) = 0$ for $x \neq y$ points that the frequency locations are different and not colliding. The Hamming cross-correlation exemplifies the orthogonal nature of the sequences. For synchronous FHMA applications, where the relative time delay stays within one hop, the cross-correlation between every sequence $C^a$, $0\leq a < q$ at $\alpha =0$ equals $0$ indicates they are orthogonal. The Hamming cross-correlation at delay $\alpha=0$ between every sequence in set $\mathcal{C}= \{ C^0, C^1,..., C^{q-1}\} $, $0 \leq u,v < q$,  $u \neq v$ is zero, $G_{uv}(0) = 0$, for every $u$ and $v$ showing that all the sequences in the set are orthogonal to each other.

\begin{figure}[h]
\centerline{\includegraphics[scale=0.165]{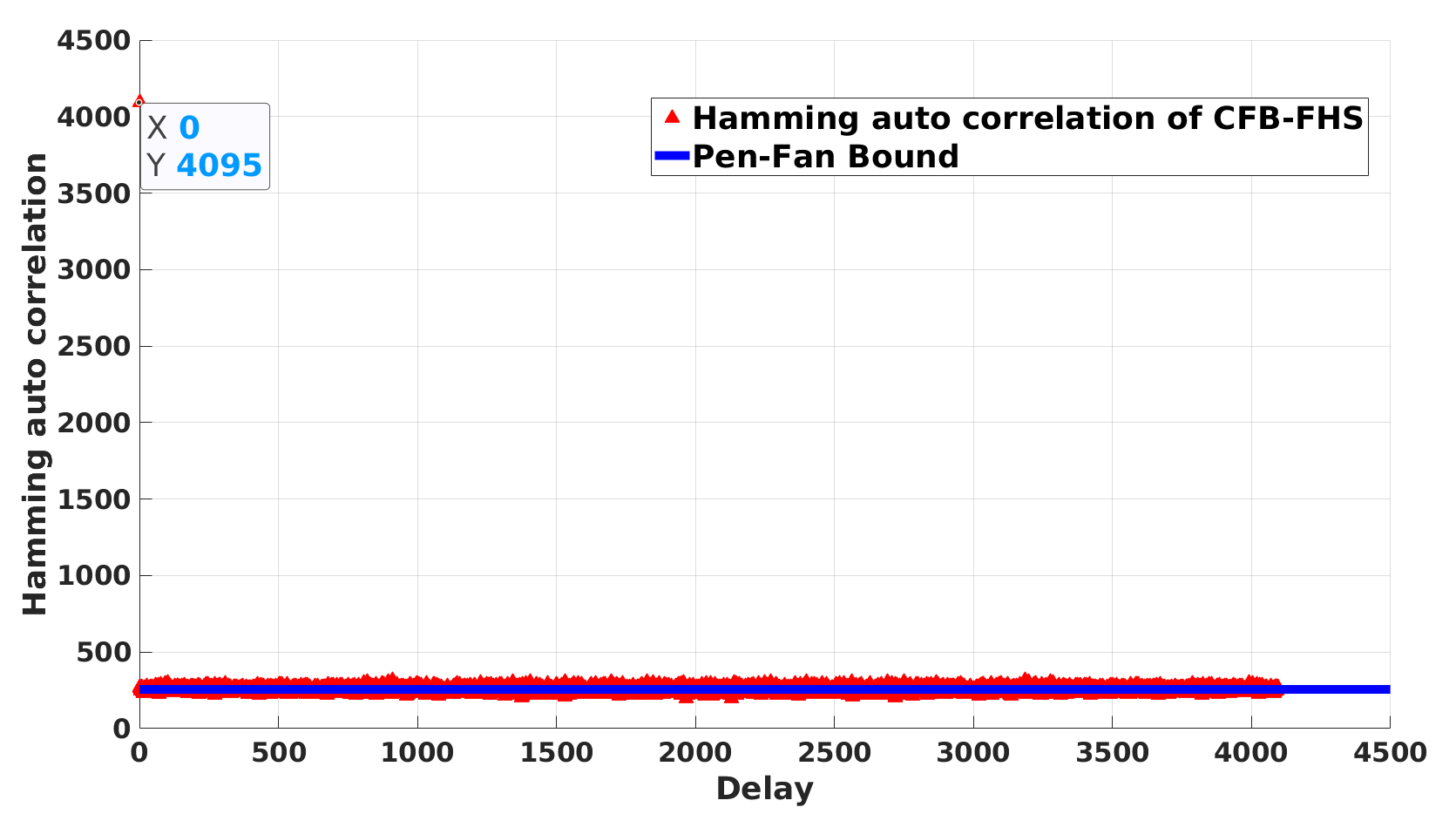}}
\caption{Hamming autocorrelation of CFB-FHS, $M = 16, q = 4, l = 14, n' = 4095$}
\label{ham_ac_with_penfan_l_14_m_16}
\end{figure}

\begin{figure}[h]
\centerline{\includegraphics[scale=0.163]{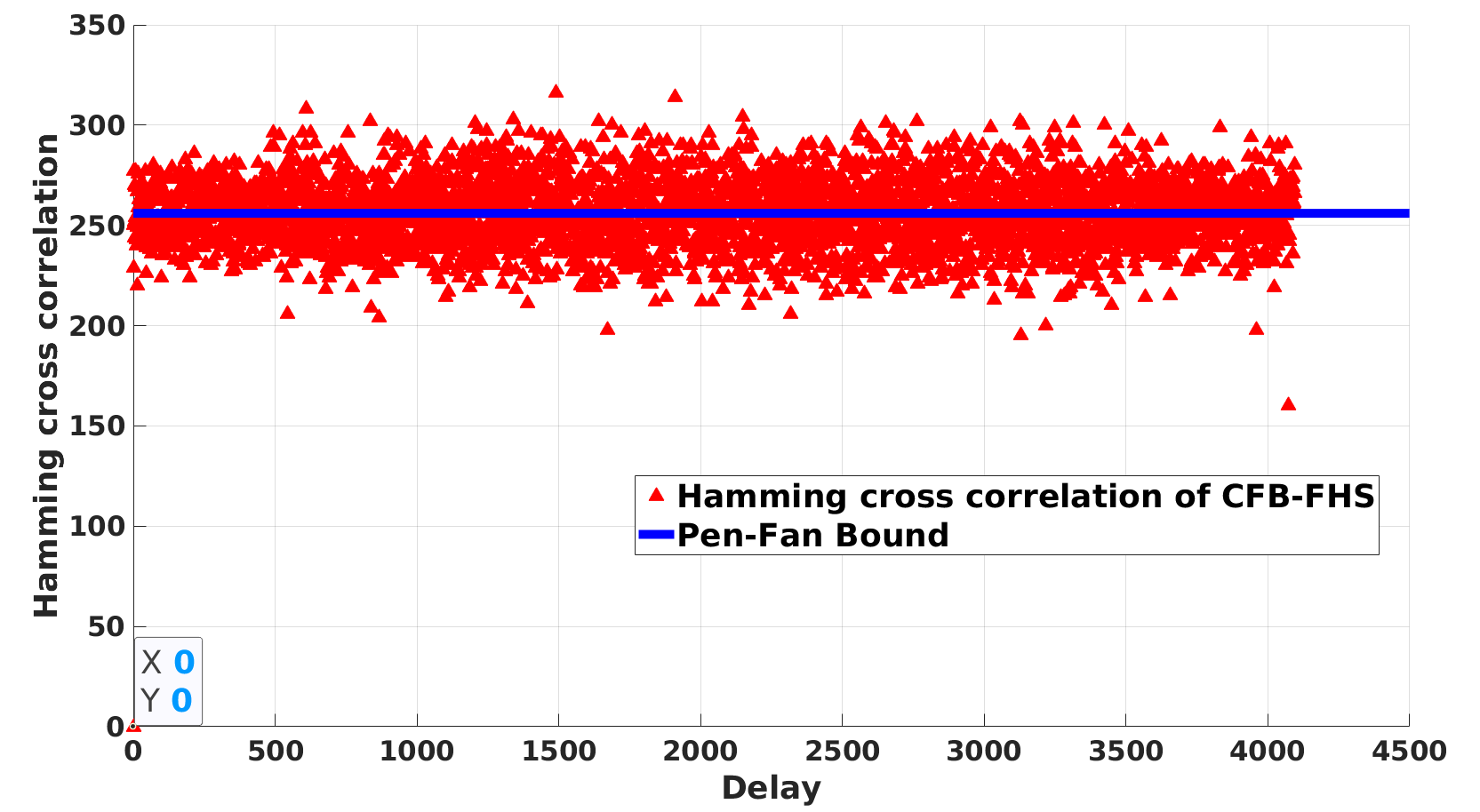}}
\caption{Hamming cross-correlation within CFB-FHS Set, $M = 16, q = 4, l = 14, n' = 4095$}
\label{ham_cc_with_penfan_l_14_m_16}

\end{figure}

In \cite{b3}, the authors derive a lower bound for maximum Hamming correlation, $G_{m}$ of any FHS set, which is given by,
\begin{equation}
G_{m} \geq \frac{(n'q - M)n'}{(n'q - 1)M}
\label{Pen-Fan bound}
\end{equation}
Any FHS set that attains the equality in this Pen-Fan bound \eqref{Pen-Fan bound} is called an optimal FHS set. Generally, the maximum Hamming cross-correlation of any sequences is dependent on the length of the sequences as well. The proposed CFB-FHS set is not an optimal FHS with regard to \eqref{Pen-Fan bound}, but has low Hamming cross-correlation values around the Pen-Fan bound, irrespective of having a very long sequence length. As an example, the Hamming auto-correlation and cross-correlation plot for sequences from our CFB-FHs set $\mathcal{C}$ along with the Pen-Fan bound \eqref{Pen-Fan bound} from\cite{b3}  is shown in Fig.~\ref{ham_ac_with_penfan_l_14_m_16} and Fig.~\ref{ham_cc_with_penfan_l_14_m_16}, respectively.

\begin{table}
\begin{center}
\caption{Comparison of the length between different FHS sets}

\begin{tabular}{|p{1.6 cm}|p{1.2 cm}|p{0.65 cm}|p{1.2 cm}|p{1.1cm}|p{0.45 cm}|}
\hline
  & \textbf{Number of frequency spots, $M$} & \textbf{Family Size, $q$} & \textbf{Degree of Polynomial, $l$} & \textbf{Length of the sequence} &\textbf{$Z_{NH}$} \\ 
  \hline 
  \multirow{2}{*}{New CFB-FHS} &\hfil 16 & \hfil 5 & \hfil 14 & 4095 & 0 \\
  \cline{2-6}
          &\hfil 16 & \hfil 15 & \hfil 14 & 4095 & 0 \\
  \hline
  \multirow{2}{*}{S-NHZ} &\hfil 16 & \hfil 5 & \hfil - & 16 & 2 \\
\cline{2-6}
   & \hfil 16 & \hfil 15 & \hfil -& 16 & 0 \\
   \hline
   \multirow{2}{*}{ML NHZ} & \hfil 16 & \hfil 5 & \hfil - & 240 & 2 \\
\cline{2-6}
   & \hfil 16 & \hfil13 &\hfil - & 210 & 0 \\
\hline
\end{tabular}
\label{length_comp}
\end{center}
\end{table}

\section{Conclusion}

This paper presented a novel algorithm for constructing an O-FHS set, which is suitable for synchronous FHMA applications and uses $m$-sequence as the base sequence. The proposed CFB-FHS construction was explained. Some of the key properties of the generated sequence set were studied with supporting results and compared with existing literature. The CFB-FHS set has a longer period and excellent correlation properties. It uses all available frequency slots equally, retains the random nature of the sequences, and can support the maximum number of simultaneous users bounded by the expression in \eqref{q constraint}. This sequence set is more suitable for modern synchronous FHMA applications where the frame sync timing offset between two FH pairs of nodes could be less than the dwell time of the FH system. Future work could include extensions of the sequences for quasi-synchronous FHMA applications, where the FHS set will remain orthogonal to each other even when the relative time delay between them is up to a few dwell periods of the FHMA system.

\end{document}